\documentclass[draft]{elsarticle}
\usepackage{color}
\usepackage{times}
\usepackage{vdm}
\usepackage{url}

\leftRecord
\leftCases
\def\DeF{\;\raise.5ex
        \hbox{\footnotesize\underline{$\mathchar"3234$}}\;}% a \triangle
\newcommand{\refsto}{\mathrel{\sqsubseteq}}

\newcommand{\Guarantee}{\mathop{\kw{guar }}}

\newcommand{\Rely}{\mathop{\kw{rely}}}
\newcommand{\rely}[2]{\Rely #1 \suchthat #2}

\newcommand{\Postcondition}{\mathop{\kw{post}}}

\newcommand{\Var}{\mathop{\kw{var}}}

\DeclareSymbolFont{largesymbols}{OMX}{yhex}{m}{n}
\DeclareMathAccent{\wideparen}{\mathord}{largesymbols}{"F3}
\newcommand{\posvals}[1]{\wideparen{#1}}
\newcommand{\evalvals}[1]{\wideparen{\wideparen{#1}}}
\newcounter{defsThms}
\newcommand{\atomicassn}[1]{\langle #1 \rangle}

\newcommand{\spot}{\mathrel{\suchthat}}

\makeatletter
\def\Spec{\@ifnextchar*{\@Spec}{\@@Spec}}
\def\@Spec*#1#2#3{\ifx\@empty#1\else#1\colon\fi
   [#2\ifx\@empty#2\else,\ \fi#3]}
\def\@@Spec#1#2#3{\ifx\@empty#1\else
   \begin{array}[b]{@{}l@{}}#1\colon\end{array}\!\fi%
   \left[\ifx\@empty#2\else\begin{array}[b]{@{}c@{}}#2\end{array},\ \fi%
   \begin{array}[b]{@{}c@{}}#3\end{array}\right]}
\makeatother

\begin{document}

\journal{Logical and Algebraic Methods in Programming}

\begin{frontmatter}

\title{Possible values:
exploring a concept for concurrency}
\author[nu]{Cliff B. Jones\corref{cor1}} % \fnref{fn1}}
%\ead{Cliff.Jones@newcastle.ac.uk}
\author[uq]{Ian J. Hayes}
%\ead{Ian.Hayes@itee.uq.edu.au}
\cortext[cor1]{Corresponding author}
%\fntext[fn1]{Footnote 1}
\address[nu]{School of Computing Science, Newcastle University, UK}
\address[uq]{School of Information Technology and Electrical Engineering, The University of Queensland, Australia}

\begin{abstract} %%%%%%%%%%%%%%%%%%%%%%%%%%%%%%%%%%%
An important issue in concurrency is interference.
This issue manifests itself in both shared-variable
and communication-based concurrency --- this paper focusses on the former case where interference is caused by the environment of a process changing the values of shared variables.
Rely/guarantee approaches have been shown to be useful in specifying and reasoning
compositionally about concurrent programs.
This paper explores the use of a ``possible values'' notation for reasoning about variables whose values can be changed multiple times by interference.
Apart from the value  of this concept in providing clear specifications,
it offers a principled way of avoiding the need for some auxiliary (or ghost) variables
whose unwise use can destroy compositionality.
%The possible values concept also helps sharpen some issues around atomicity.
\end{abstract}


\end{frontmatter}

% version!
\newcommand{\version}{3.5 (for journal)}

% comment out to restore standard runningheads
%\newcommand{\kopf}{\textnormal{\today.~Version~\version}}
%\pagestyle{myheadings}
%\markboth{\kopf}{\kopf}
%
%\fbox{Version \version}\hfill
%\fbox{Dated: \today}

\section{Introduction} %%%%%%%%%%%%%%%%%%%%%%%%%%%%%
\label{S-intro}

High on the list of issues that make the design of concurrent programs difficult to get right is `interference'.
Reproducing a situation that exhibited a `bug' can be frustrating;
attempting to reason informally about all possible interleavings of interference can be exasperating;
and designing formal approaches to the verification of concurrent programs is challenging.

Recording post conditions for sequential programs applies the only real tool that we have:
abstraction is achieved by winnowing out what is inessential in the relationship between the initial and final states of a computation.
Post conditions record the required relationship without fixing an algorithm to bring about the transformation;
furthermore, they record required properties only of those variables which the environment will use.
The rely/guarantee approach (see~Section~\ref{S-RG})
uses abstraction in the same way to provide specifications of concurrent software components that are more abstract than their implementations:
for any component,
rely conditions are relations that record interference that the component must tolerate and 
guarantee conditions document the interference that the environment of the component must accept.

This paper explores a concept that fits well with rely/guarantee reasoning but probably has wider applicability.
In relational post conditions, it is necessary to be able to refer to the initial value $x$ and final value $x'$ of a variable $x$
(e.g.~$x \leq x' \leq x + 9$).
If however it is necessary to record something as simple as the fact that a local variable $x$
captures one of the values of a shared variable $y$,
it is inadequate to write $x' = y \Or x' = y'$ in the case where $y$ might be changed many times by the environment.
Enter `possible values':
the suggested notation is that $\posvals{y}$ denotes the set of values which variable $y$  contains during the execution of the operation in whose specification $\posvals{y}$ is written. So,
(assuming the access to read the value of $y$ is atomic):

\begin{formula}
post-Op: x' \in \posvals{y}
\end{formula}

\noindent
is satisfied by a simple assignment of $y$ to $x$.

\subsection{Rely/Guarantee thinking}
\label{S-RG}

Before going into more detail on the possible values notation
(see~Section~\ref{S-pv}),
a brief overview of background work is offered.
The specifications given in Section~\ref{S-4slot} are written 
in the notation of VDM~\cite{Jones80a,Jones90a}.
It is unlikely that they will present difficulties even to readers unfamiliar with that specific notation because similar ideas for sequential programs are present in
Z~\cite{Hayes93},
B~\cite{Abrial96},
Event-B~\cite{Abrial10},
and
TLA~\cite{lamport03}.
The basic idea is of state-based specifications with operations (or events) 
transforming the state and  being specified by something like pre and post conditions.
Pre conditions are predicates over states that indicate what can be assumed about states in which an operation can be initiated.
Post conditions are relations over initial and final states that specify the required relations between the initial and final values of state components.
Good sequential specifications eschew any details of implementation algorithms:
they do not specify anything about intermediate states;
in fact an implementation might use a state with more components.
At first sight, it might appear surprising that there is not a precise functional requirement on the final state but using
non-determinism in specifications turns out to be an extremely useful way of postponing design decisions.

The use of abstract objects in specifications is a crucial tool for larger applications.
Moreover, datatype invariants can make specifications clearer:
restricting types by predicates simplifies pre/post conditions and
also offers a way for the specifier to record the intention of a specification.
Another useful aspect of VDM is the ability to define more tightly the `frame' of an operation by recording whether access to state components is for (only) reading or for both reading and writing.%
\footnote{Most of the literature on rely/guarantee conditions
is limited to normal (or `scoped') variables;~\cite{SEFM-15-paper} shows how `heap' variables can be viewed as representations of more abstract states.}

The basic rely/guarantee~\cite{Jones81d,Jones83a} idea%
\footnote{The literature on rely/guarantee approaches continues to expand;
see~\cite{FACJexSEFM-14,HayesJonesColvin14TR}  for further references.
For a reader who is completely unfamiliar with rely/guarantee concepts,
a useful brief presentation can be found in~\cite{Jones96a}.}
is simple:
interference is documented and proof rules are given which support reasoning about interference in 
concurrent threads.
Just as in sequential specifications,
the role of a state is central to recording rely/guarantee specifications.
For concurrency,
it is accepted that the environment of a process can change values in the state during execution of an operation.%
\footnote{Notice that there is an essential difference here from `actions'~\cite{BackAtomicityRefinement}
or `events'~\cite{Abrial10} which view execution of a guarded action as atomic.}
Such changes are however assumed to be constrained by a rely condition.
In order to reason about the combined effect of operations,
the interference that a process can inflict on its environment is also recorded;
this is done in a guarantee condition.
Both rely and guarantee conditions are, for obvious reasons,
relations over states.
In the original form 
--and after many experiments--
 both conditions are reflexive and transitive
covering the possibility of zero or many steps.
Such relations often indicate monotonic evolution of variables.

It is useful to compare the roles of rely and guarantee conditions with the better known pre/post conditions.
Pre conditions are essentially an invitation to the designer of a specified component to ignore some starting states;
in the same way, the developer can ignore the possibility that interference will make state changes that do not satisfy the rely condition.
In neither case should a developer include code to test these assumptions;
there is an implicit requirement to prove that the component is only used in an appropriate context.
In contrast, post conditions and guarantee conditions are obligations on the running code that the developer has to create;
these conditions record properties on which the deployer can depend.

\begin{figure}
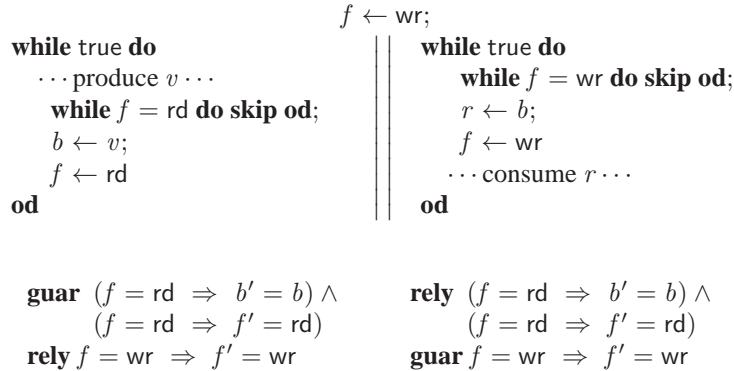

\begin{center}
\begin{eqnarray*}
\begin{array}{c}
f \gets \mathsf{wr};\\
 \left.
  \begin{array}{l}
   \kw{while}\ \mathsf{true}\ \kw{do}\\
    \ \ \begin{array}{l}
     \cdots \mbox{produce $v$} \cdots\\
     \begin{array}{l}
     \kw{while}\ f=\mathsf{rd}\ \kw{do skip od};\\
     b \gets v;\\
     f \gets \mathsf{rd}
     \end{array}\\
  \end{array}\\
   \kw{od}
  \end{array}
 \ \ \right|\left|\ \ 
 \begin{array}{l}
  \kw{while}\ \mathsf{true}\ \kw{do}\\
   \ \ \begin{array}{l}
     \begin{array}{l}
     \kw{while}\ f=\mathsf{wr}\ \kw{do skip od};\\
     r \gets b;\\
     f \gets \mathsf{wr}
     \end{array}\\
    \cdots \mbox{consume $r$} \cdots
   \end{array}\\
   \kw{od}
  \end{array} 
  \right.
\end{array}
\end{eqnarray*}
%%%%%%%%%%%%%%%%
\begin{eqnarray*}
 \left.
  \begin{array}{l}
   \kw{guar} 
     \begin{array}[t]{l} 
       (f=\mathsf{rd} \Implies b'=b) \And {} \\
       (f=\mathsf{rd} \Implies f'=\mathsf{rd})
     \end{array} \\
   \kw{rely } f=\mathsf{wr} \Implies f'=\mathsf{wr}
 \end{array}
 \ \
 \begin{array}{l}
  \begin{array}{l}
   \kw{rely} 
     \begin{array}[t]{l}
       (f=\mathsf{rd} \Implies b'=b) \And {} \\
       (f=\mathsf{rd} \Implies f'=\mathsf{rd})
     \end{array}\\
   \kw{guar } f=\mathsf{wr} \Implies f'=\mathsf{wr}
 \end{array}
  \end{array} 
  \right.
\end{eqnarray*}\end{center}
\caption{A one-place buffer}
\label{F-1place}
\end{figure}

The simplest form of relation that could be used in rely or guarantee conditions is to state that the value of a variable remains unchanged (e.g.~$b' = b$).
Such unconditional constraints are normally better handled by marking an operation
(or part thereof) as having only read access.
There is however an important way to combine `monotonic' changes to flags with assertions about variables remaining unchanged.
Consider a simple one-place buffer 
in which a producer process places a value in a buffer variable $b$ from which a consumer process extracts values. 
Testing and setting flag $f$ in Figure~\ref{F-1place}
ensures that the producer and consumer alternate their access to $b$.
During its read phase, the consumer needs to rely on the fact that the value of $b$ cannot change but
this is too strong as a rely condition for the whole of the consumer process ---
the producer process could never insert anything into the buffer if it were required to achieve a guarantee condition of $b' = b$.
But the consumer process can instead rely on $f = \mathsf{rd} \Implies b' = b$, which in turn is easy for the 
producer to guarantee. 
The `monotonic' behaviour of the flags means that the producer has also to guarantee that $f = \mathsf{rd} \Implies f' = \mathsf{rd}$
and the consumer must guarantee $f = \mathsf{wr} \Implies f' = \mathsf{wr}$.
This example shows one way in which rely/guarantee conditions can be used to reason about race-free programs.
It also illustrates a technique that is used in Section~\ref{S-4slot} to locate what is going on in the environment without adding auxiliary variables.
The example tackled in Section~\ref{S-4slot} is however much more challenging than this simple one-place buffer.

\subsection{Law for mutual strengthening of guarantee and rely}\label{S-strengthening-law} %%%%%%%%%%%%%%%%%%%%%%%%%%%%%

As part of the example in Section~\ref{S-4slot}, 
a new facet of rely/guarantee refinement is needed:
it allows mutual strengthening of both rely and guarantee conditions 
for a portion of one process.
The approach is a contribution to rely/guarantee refinement and 
%while it does not make use of possible values concepts,
it makes it possible to avoid introducing additional auxiliary variables 
(see~Section~\ref{s:auxiliary})
in order to handle the example in Section~\ref{S-4slot}.

In the standard approach to rely/guarantee refinement,
when two parallel processes are introduced each has an associated rely/guarantee pair
and there is an obligation to show that the guarantee of each implies the rely of the other.
Normally the one rely/guarantee pair suffices to handle the refinement of a process
but for the example in Section~\ref{S-4slot} that is not sufficient.

In the standard theory, rely/guarantee pairs are often mutually dependent:
for the two-process case,
a process $P$ maintaining its guarantee may be dependent on its environment
(process $Q$) 
maintaining the rely of $P$ (by $Q$ maintaining its guarantee)
and vice versa.
For example, $P$ may guarantee to maintain $x \geq 0$
provided it can rely on its environment maintaining $x \geq 0$.
The guarantee, $g$, of a process has to hold for every atomic program step it makes
and hence $g$ has to be weak enough to be maintained by every step.
However, for a subpart $S$ of $P$, 
all the atomic steps of $S$ may imply a stronger guarantee $gs$.
As $P$ forms the environment of process $Q$, while $P$ is executing subpart $S$,
$Q$ may assume a stronger rely condition of $gs$ and as a consequence of this 
its own guarantee may be strengthened from $r$ to $rs$,
which in turn allows process $P$ to assume a stronger rely condition $rs$,
but only while it is executing subpart $S$.
Note that while only a subpart $S$ of $P$ is of concern,
the whole of $Q$ has to be considered for the strengthening of its rely and guarantee.

In order to establish the strengthened rely/guarantee pair for the duration of $S$,
the state when $P$ enters $S$ may need to satisfy an initial condition $j$.
%(which in some cases forms an invariant for the execution of $S$). 
For the example in Section~\ref{S-4slot} a special case of the above reasoning
applies in which the guarantee of $P$ is strengthened to state that 
$P$ does not modify any shared variables.
In this case one needs to show that process $Q$ maintains the stronger rely $rs$
from any initial state satisfying $j$ provided $Q$ suffers no inference from $P$.

\subsection{Connection to data abstraction/reification}\label{S-data-reification} %%%%%%%%%%%%%%%%%%%

It  is important to appreciate how rely relations abstract from the detail of the actual environmental interference of an operation. 
Obviously, the most detailed information about an environment is the actual state changes it makes.
But designing to such concrete detail would create a component that is not robust to change.
Just as post conditions deliberately omit implementation details of a specified operation,
it is useful to strive for an abstract documentation of interference.
It is clear that relations cannot record certain sorts of information but, if they are adequate for a given task, their use will yield a more compositional development than the detail of the environment.

The extended example in Section~\ref{S-4slot} shows the importance of linking rely/guarantee ideas with data abstraction and reification.
Specification using abstract mathematical objects and the process of stepwise introduction of more concrete
(i.e.~closer to hardware)
objects is well established for sequential programs and for significant applications is often more telling than the abstraction that comes from post conditions ---
see, for example,~\cite{Jones90a}.
In addition to layering design decisions, 
careful use of abstract objects in the development of concurrent programs 
offers other advantages.
In particular, 
developments can appear to allow data races at an abstract level that are removed by careful choice of a concrete representation ---
this is discussed in~\cite{Jones06a}.
One reason that this is interesting is Peter O'Hearn's suggested dichotomy in~\cite{OHearn07}
that separation logic is appropriate 
for reasoning about race avoidance whilst rely/guarantee methods fit `racy' programs.
The distinction between abstract and concrete data races is perfectly illustrated in Section~\ref{S-4slot} but the example is not easy to summarise.
A  simpler example is searching an array to find the lowest index of an element that satisfies a predicate $P$
by means of two parallel processes that search the elements with, respectively, even and odd indices
(for a full development of this example, see~\cite{HayesJonesColvin14TR}).
If a single variable $t$ were used to record the least index of an element that satisfies $P$,
it would be necessary to have locks in the the two processes 
to avoid a data race on $t$.
A neat way to avoid the `write/write' race is to represent $t$ by the minimum of two variables, $et$ and $ot$ 
that record the least value of, respectively, even and odd indices where the array element satisfies $P$.
The `write/write' race, which is useful in an abstract description of the design, is reduced to a `read/write' race because the actual code for each process updates only one of the variables
although it reads the other variable in its loop test
(and on the completion of both processes $t$ can be retrieved as $min(et,ot)$).

The citations above relate to the original form of rely/guarantee reasoning in which the (potentially) four conditions are combined.
More recent work
has shown how separate rely and/or guarantee constraints can be wrapped around any command including conventional refinement calculus style specifications.
The presentation in~\cite{FACJexSEFM-14,HayesJonesColvin14TR} of rely/guarantee thinking 
makes algebraic properties clearer.

\subsection{Plan of this paper}

This paper  provides evidence of the usefulness of the possible values concept.
Section~\ref{S-pv} presents a notation for the concept
while Section~\ref{S-4slot} is an extended  example using the concept and notation.
Section~\ref{S-sem} outlines how a semantic model can be provided and looks at the form of laws that would fit the newer presentation of rely/guarantee 
reasoning~\cite{HayesJonesColvin14TR,FACJexSEFM-14}.
The current authors recognise that this paper  represents the start of an exploration ---
some avenues to be investigated are mentioned in Section~\ref{S-conc}.
%They hope that the exploratory nature of the contribution will please the dedicatee!

\section{Possible values} %%%%%%%%%%%%%%%%%%%%%%%%%%%%%
\label{S-pv}

It is argued above that the confessed expressive weakness of rely/guarantee specifications serves the purpose of preserving some form of compositionality in the
design of concurrent programs.
However,
if notations can be found that increase expressive power,
they should be evaluated both for expressiveness and tractability.
The simple case mentioned above of using one or more possible values terms in a post condition is considered first and issues about extension are deferred to Section~\ref{S-conc}.

\subsection{Possible values of variables} %%%%%%%%%%%%%%%%%%%%%%%%%%%%%

If an operation only has read access to a shared variable $y$
and $x$ is a local variable of the process, then:
\begin{eqnarray}\label{posvals-example1}
post-Op: x' \in \posvals{y}
\end{eqnarray}
\noindent
requires that the final value of the variable $x$ should contain one of the values that the environment places in the variable $y$ ---
this includes the (initial) value of $y$ at the time $Op$ began execution.
So $\posvals{y}$ denotes a set of values whose elements have the type of $y$.

Notice that the post condition above is `stable' in the sense that the environment might change the value of $y$ after $Op$ accesses the variable and the post condition is still true.
In contrast, it would be unwise to write a post condition that contained
$x' \notin \posvals{y}$
because this would not be stable and it would appear to require that every possible change that the environment makes to the value of $y$ is observed.
(In some cases, it would be possible to establish such a result 
under a suitable rely condition;
but some form of (local) datatype invariant should also be considered in such cases.)

So, for the straightforward case,
the post condition (\ref{posvals-example1}) can be established by the atomic assignment $x \gets y$.
As is reported in Section~\ref{S-ACM-approaches},
an instance of this simple case was the inspiration for the possible values notation.
There are however several vectors of extension.
If the process in which the $\posvals{y}$ term is written also has write access to the variable $y$,
it is necessary to take a position on whether both environment assignments to $y$ and those of the component itself are reflected in $\posvals{y}$;
the view of the current authors is that $\posvals{y}$ contains 
all values of $y$ that could be observed by the process.

\subsection{Semantics and laws}\label{S-laws} %%%%%%%%%%%%%%%%%%%%%%%%%%%%%
\label{S-sem}

%\cliffnote{This section needs re-visiting once Section~\ref{S-1step} is worked out}

It is not difficult to see how a formal meaning can be given to the simple form of 
the possible values notation in a semantics such as that in~\cite{HayesJonesColvin14TR}:
%Ian: Im completely happy to change this to the CONCUR paper if you concur
basically, that portion of the sequence of states that corresponds to the execution of an operation 
is distinguished so as to identify the first and last states in order to give a semantics to post conditions.
It is only necessary to consider all of the states in that portion and to extract the set of values 
of the relevant variable.
%This simple case covers the usage of possible values in the example of Section~\ref{S-4slot} 
%and all of the others which have been driven by practical applications.
%
%
%\iannote{Omit? This contradicts the end of the previous section.\\
%A decision has to be made about the required meaning in the case
%where both the process 
%(in whose post condition the possible values term is used)
%and its environment can change the relevant variable.
%This topic will be addressed when a practical application shows the need 
%(and hopefully gives a steer as to the most appropriate choice).
%}

Another interesting semantic issue concerns locking.
In fact, the possible values notation forces consideration of a number of facets of `atomicity'.
%\subsection{Locking}\label{s:locking} %%%%%%%%%%%%%%%%%%%%%%%%%%%%%%%%
Locking may be used to ensure mutually exclusive access to a set of variables.
A process may lock a resource protecting a set of variables.
While it owns the lock, it may make multiple changes to the variables protected by the lock,
however, any other processes accessing the protected variables cannot observe 
any of the intermediate states of the protected variables.
Hence a process in the scope of a resource with a set of protected variables can only 
observe the initial and final states of a protected block within another process.
Throughout the body of a protected block a process can rely on the protected variables being stable.
Furthermore, any guarantee involving just the protected variables has to hold only between
the initial and final states of the protected block.

Just as the semantics for the straightforward use of possible values terms in a post condition
poses no difficulties in terms of the underlying traces,
a rather simple law suffices to reason about the notation.
Here, it is convenient to switch to the refinement calculus style 
of~\cite{FACJexSEFM-14,HayesJonesColvin14TR}
in which the specification statement $\Spec{x}{}{q}$ establishes the postcondition $q$
and modifies only $x$, 
and the command $c$ in a rely context of $r$ is written $\rely{r}{c}$.
Assuming a read of $y$ is atomic, the following law holds.

\begin{eqnarray}\label{law-posvals-y}
  \rely{(x' = x)}{\Spec{x}{}{ x' \in \posvals{y}}} \ \ \ \refsto\ \ \  x \gets y
\end{eqnarray}

\noindent
The rely condition $x' = x$ is required to ensure that the environment doesn't change $x$
after the assignment is made.
For example, $x$ may be a local variable or,
as below in Section~\ref{S-4-spec}, annotated $\kw{owns wr}\ x$.

\subsection{Possible values of expressions} %%%%%%%%%%%%%%%%%%%%%%%%%%%%%

For the set of possible values of an expression, $\posvals{e}$, 
one needs to consider the corresponding set of states of the execution
and form the set of values of $e$, each evaluated in one of those states.
Importantly, all values of program variables used in $e$ are sampled in a single
state for each evaluation.
For example, for the specification
\begin{eqnarray}\label{posvals-y-plus-y}
  \Spec{x}{}{ x' \in \posvals{y+y} }
\end{eqnarray}
both occurrences of $y$ are sampled in the same state
and hence the resultant value is always even
(assuming the variables are integer valued).
Note that there is a subtle difference between (\ref{posvals-y-plus-y}),
which samples $y$ once, and
\[
  \Spec{x}{}{ \exists{v,w}{v \in \posvals{y} \And w \in \posvals{y} \And x' = v+w} }
\]
which samples $y$ twice so that the values of $v$ and $w$ may differ.

Replacing $y$ in the law~(\ref{law-posvals-y}) with an expression $e$ introduces the complication that 
each variable reference in the evaluation of $e$ in the assignment could be accessed in a different state.
Note that if $e$ has multiple references to a single variable $y$,
each reference could be accessed in a different state.
However, if $e$ has only a single reference to a variable $y$ and 
all other variables in $e$ are stable, any evaluation of $e$ is equivalent to evaluating it 
in the state in which $y$ is accessed and the law is valid.
Let $S$ be a set of variables such that the free variables of $e$ are contained in $S \union \{ y \}$
and $e$ has only a single reference to $y$ and accesses to $y$ are atomic, then
\begin{eqnarray}\label{law-single-reference-posvals}
  \rely{(x' = x \And (\bigwedge z \in S \spot z' = z))}{\Spec{x}{}{x' \in \posvals{e}}} \ \ & \refsto & \ \  x \gets e\ .
\end{eqnarray}

If $e$ is of the form $d(f)$ for a mapping $d$ and expression $f$,
stability is required on the program variables in $f$
but stability is not required for the whole of $d$, just $d(f)$,
because the other elements of $d$ have no effect on the expressions value.

If the expression $e$ contains multiple references to a variable $x$,
saving $x$ in a local variable $t$ and then evaluating $e$ in terms of $t$ 
ensures that the value used for $x$ is from a single state.
The following refinement law ensures $x$ is sampled once.
It is assumed that $r$ and $t$ are local variables 
(and hence the environment cannot change them)
and that $r$ and $t$ do not occur free in $e$. 
%but that $v$ does occur free in $e$
%(although the law still holds without this assumption but isn't much use).
\begin{eqnarray}\label{law-posvals-pre-assignment}
  \Spec{t,r}{}{ r' \in \posvals{e[x/v]} } 
  \ \ \refsto\ \ 
  \atomicassn{t \gets x} \ ;\  \Spec{r}{}{ r' \in \posvals{e[t/v]} }
\end{eqnarray}
For this to be valid one needs to rely on the environment maintaining 
$\posvals{e[t/v]} \subseteq \posvals{e[x/v]}$,
for the duration of the command.
This holds provided the rely condition
\[
  x' \neq t \land t' = t \implies e[t/v]' = e[t/v]
\]
is maintained by the environment,
where $e[t/v]'$ stands for $e[t/v]$ evaluated in the after state,
i.e.\ $e'$ is $e$ with every program variable $y$ in $e$ replaced by $y'$.

Law (\ref{law-posvals-pre-assignment}) can be justified as follows.
The atomic statement $\atomicassn{t \gets x}$ establishes $e[t/v] = e[x/v]$.
An environment step that has a final state in which $x' = t$ establishes 
$e[t/v]' = e[x/v]'$
otherwise the environment establishes
$e[t/v]' = e[t/v] = e[x/v]$.

As an example consider the case in which the expression $e$ is $d(v)$.
Applying (\ref{law-posvals-pre-assignment}) gives
\begin{eqnarray}
  \Spec{t,r}{}{ r' \in \posvals{d(x)} } 
  \ \ \refsto\ \ 
  \atomicassn{t \gets x} \ ;\  \Spec{r}{}{ r' \in \posvals{d(t)} }
\end{eqnarray}
provided its environment ensures the condition: $x' \neq t \implies d'(t) = d(t)$.
Immediately after the atomic assignment to $t$, 
\begin{equation}\label{posvals-d(t)}
  d(t) = d(x) \in \posvals{d(x)}\ .
\end{equation}
If the environment makes a step that does not change $x$, 
(\ref{posvals-d(t)}) is maintained because $d'(t') = d'(x')$
but if the environment changes $x$ so that it no longer equals $t$
one can no longer rely on $d'(t')$ being the same as $d'(x')$.
However, if one can rely on $d(t)$ being stable
and because $d(t) = d(x)$ and $d'(t') = d(t)$,
one can still deduce $d'(t') = d(x)$.

\section{Asynchronous Communication Mechanisms} %%%%%%%%%%%%%%%%%%%%%%%%%%%%%
\label{S-4slot}

An Asynchronous Communication Mechanism (ACM)
logically provides a one-place buffer between a single writer and a single reader
(see Figure~\ref{F-loops}).
This sounds trivial but the snag is in the adjective:
ACMs are asynchronous in the sense that neither the reader nor the writer 
should ever be held up by locks.%
\footnote{\label{FN-cf}This contrasts with the simple one-place buffer in Section~\ref{S-intro}
where the code would `busy wait' on the value of a flag to control alternation between the producer and consumer.}
Unless the value being communicated via the buffer is small enough to be read and written atomically,
it should be obvious that one slot is not enough to realise the buffer; 
a little thought shows that a buffer representation with two slots is also inadequate;
the topic of how many slots are required is returned to in Section~\ref{S-reif1}.
In~\cite{Simpson90}, % Simpson97}
Hugo Simpson proposed a `four-slot' algorithm 
to implement an ACM for which,
while the code is short,
extremely subtle reasoning is required for its justification.

\begin{figure}
\begin{center}
\begin{eqnarray*}
 \left.
  \begin{array}{l}
   \kw{while } \mathsf{true} \kw{ do}\\
    \ \ \begin{array}{l}
     \cdots \mbox{produce $v$} \cdots\\
     Write(v)
    \end{array}\\
   \kw{od}
  \end{array}
 \ \ \right|\left|\ \ 
 \begin{array}{l}
  \kw{while } \mathsf{true} \kw{ do}\\
   \ \ \begin{array}{l}
    r \; \gets \; Read()\\
    \cdots \mbox{consume $r$} \cdots
   \end{array}\\
   \kw{od}
  \end{array} 
  \right.
\end{eqnarray*}
\end{center}

\caption{Code to clarify reader/writer structure}
\label{F-loops}
\end{figure}

\subsection{ACM requirements}\label{S-ACM-requirements}

The requirement is to 
communicate the ``most recent'' value from a single producer to a single consumer via a shared buffer.
More precisely, it must satisfy the following.
  \begin{itemize}
  \item It is assumed that there is only a single reader and a single writer but the reader and writer processes operate completely asynchronously
  \item A write puts a new value in the buffer
  \item A read gets a completely written value from the buffer
  \item The value read is at least as fresh as the last completely written value when the read started -- 
  this implies that, for two consecutive reads, the value read by the second read will be at least as fresh as
  that read by the first
  \item Reads and writes must not block (no locks)
  \item Reads and writes of values can't be assumed to be atomic (i.e.\ a single value may be larger than the atomic changes made by the hardware)
  \item The only thing Simpson assumes to be atomic is the setting of single bits (and they are actually realised by wires)
  \item The buffer is initialised with a data value (so there is always something to read)
  \item The buffer is shared by the reading and writing processes alone (i.e.\ no third process can modify the buffer)
  \end{itemize}

%\draftnote{
%The referee says that in the terminology of Lamport \cite{Lamport86II} this can be summarised as
%``implementing a single-reader wait-free atomic register''
%but Lamport's paper does not use the term wait-free as far as I can see.
%Lamport does distinguish between safe, regular or atomic registers.
%He also distinguishes between boolean or multi-valued registers and
%he distinguishes between single reader or multi-reader.
%I need to investigate further. Ian
%}
In the terminology of Lamport \cite{Lamport86II} this can be summarised as
implementing a single-reader wait-free atomic register in terms of atomic Boolean control registers.

\subsection{Approaches to specifying ACM}\label{S-ACM-approaches}

There is an interesting range of approaches as to how the requirements that are listed above can be expressed in a formal specification.
Without surveying all of them,
it fits the theme of this paper to review two strands of publications:%
\footnote{Other approaches include~\cite{HendersonPhD,Abrial10}.}
one motivated by (Concurrent) Separation Logic~\cite{Reynolds02,OHearn07} and the other  by rely/guarantee methods.
Surveying the latter also pinpoints the origin of the possible value notation.

Richard Bornat is an expert on separation logic so it is interesting to look at how he has formalised the specification and development of Simpson's `four slot' algorithm.
In~\cite{BornatAmjad10}, separation logic is certainly used but it is interesting to see that the paper also uses rely/guarantee concepts.
In contrast,~\cite{BornatAmjad13} makes no real use of separation logic and 
the specification uses the concept of linearisability~\cite{HW90}.
The reason that this history is enlightening is that the essence of Simpson's algorithm is the exchange of `ownership' of the four slots between the reader and writer processes.
This is done precisely to ensure (data) race freedom so one would anticipate that separation logic would be in its element.
There is, in fact, one paper that uses separation logic for precisely this form of argument;
unfortunately~\cite{WangWang-10} does not include an argument that the reader always gets the `freshest' value and a recent private correspondence with one of the authors indicates that they have not extended their work to cover this essential property.

It is only fair to make an equally critical assessment of two papers~\cite{JonesPierce08,JonesPierce10} that use rely/guarantee ideas.
In the development recorded in~\cite{JonesPierce08},%
\footnote{The variable names in the Jones/Pierce papers are
$hold-r/fresh-w$;
for the reader's convenience,
these have been changed in the extracts in the current paper to match the names used here
($cr/lw$).}
it is necessary to assert that the value of one variable ($lw$)  
is assigned to another variable ($cr$);
this assertion was recorded as:

\begin{formula}
cr' = lw \Or cr' = lw'\ .
\end{formula}

\noindent
This plausible attempt says that the final value of $cr$ is either 
the initial or final value of $lw$.
Unfortunately, during the operation being specified,
the value of $lw$ could potentially be changed more than once.
This observation was precisely the stimulus that led to the invention of the notation 
for possible values.
In addition to various improvements and clarifications in the development,
the journal version~\cite{JonesPierce10} resolves the problem by using

\begin{formula}
cr' \in \posvals{lw}\ .
\end{formula}
Rushby~\cite{RushbySimpsons02} noted a similar issue in model checking Simpson's algorithm:
a version checking for just the before or after values fails in the case of multiple writes overlapping a single read.
To handle this in the model checking context, 
Rushby restricts the sequence of data values written so that they are strictly increasing in value,
and then checks that the sequence of values read is nondecreasing,
which he concludes is necessary but may not be sufficient.
He concedes that this is a limitation of the expressiveness of the model checking specification language
(which does not have the (unbounded) expressive power of the possible values notation).

There is however a deeper objection to both of the Jones/Pierce specifications of ACMs.
In both cases, the most abstract specification uses a variable ($data-w$) that contains the entire history 
of values written by the write process.
This is in spite of the fact that 
a read operation cannot access values in the sequence earlier than the last value added before the read began.
This sort of redundancy is deprecated in~\cite[Sect.~9.3]{Jones90a} as using a `biased' representation:
the state contains values that have no influence on subsequent operations.
Where there is no bias in the representation underlying a specification,
a homomorphism (retrieve function) relates a representation back to the abstraction;
in the case of a biased representation, a relation between the abstraction and the representation
is used to argue that the operations on the latter fit those on the former.
In situations where it is necessary to express non-determinism in a specification that can be removed in the design process, 
biased specifications are sometimes unavoidable ---
but, where there is an alternative, unbiased specifications should normally be preferred because they make it easier to see the range of possible implementations.
One further surprising fact about the specifications in~\cite{JonesPierce08,JonesPierce10} 
is that, even at the most abstract level, 
the specifications of both $Read$ and $Write$ are each split into two sub-operations which are joined by sequential composition.
Although the semantics of such a specification are clear,
it means that the task of convincing users that their requirements have been adequately captured
involves a rather algorithmic discussion.

Having been self-critical of these specifications, 
there is one important positive point that needs preserving in the approach below:
the issue of data-race freedom is handled in~\cite{JonesPierce10} at the level
of an abstract intermediate representation.
This is an important general point: rely/guarantee conditions can be used to record interference on an abstraction where the final code is certainly not `racy'.

\subsection{Specification using possible values}
\label{S-4-spec}
 
In contrast to the above attempts, 
a top-level specification using `possible values' notation appears to be much more natural and perspicuous.
The abstract specification uses a state with just a single value buffer $b$ of type $Value$.
The use of this intuitively simple state is only made possible by employing the possible value notation in the post condition of $Read$, 
where $\posvals{b}$ stands for the set of possible values of $b$ during the execution of $Read$.
%As in earlier specifications (e.g.~\cite{JonesPierce10}),
The $Read$ operation is described as returning a value ($r$) 
so the post condition is simply $r' \in \posvals{b}$.
This means that a single read operation can return 
the value of the write most recently completed at the time the read begins 
or of any write that executes an assignment to $b$ during the execution of the read operation.
Notice that there is no danger of a subsequent read operation obtaining an older value than the current  read because the reference point for the possible values of the newer read is the start of its execution.

As in~\cite{JonesPierce10},
the specification can be made clearer by annotating whether the external state variables accessed 
by an operation can be only read ($\kw{rd}$) 
or both read and written ($\kw{wr}$).

Thus, the specification of $Read$ can be given simply as:
    
\begin{formula}
   Read()\ r : Value \\
   \kw{ext rd } b: Value\\
%    \Guarantee b' = b \\ %    \ExtRd b; \\
    \Postcondition r' \in \posvals{b} 
\end{formula}
 
\noindent
When generating proof obligations,
the $\kw{ext rd}$ is equivalent to a guarantee condition $b' = b$.

The specification of the $Write$ operation is interesting.
If the parameter to $Write$
%(again separate from the state ($b$))
is $v$,
one would expect the post condition to be $b' = v$ ---
and this is certainly required.
In addition,
it is necessary to rule out the possibility that $Write(v)$ puts some spurious value(s) into $b$ that might be accessed by $Read$ before the $Write(v)$ corrects its wayward behaviour and achieves its post condition.
This can be expressed in a guarantee condition $b' \neq b \implies b' = v$.
Extending 
(again, as in~\cite{JonesPierce10})
the $\kw{ext}$ annotation to mark write ownership yields a specification:

\begin{formula}
   Write(v : Value) \\
   \kw{ext owns wr } b: Value\\
    \Guarantee  b' \neq b \implies b' = v \\
    \Postcondition b' = v 
\end{formula}
  
 \noindent
Here,
the proof obligation expansion of $\kw{ext owns wr}$ is
a rely condition $b' = b$,
which matches the implicit guarantee of $Read$ courtesy of its $\kw{ext rd}$ annotation.
 
The role of the guarantee of $Write$ here is to provide an intuitive specification;
the  more standard use is to show that processes can co-exist and
this usage occurs in the development below.
The guarantee of $Write$ ensures that only valid values are observable in the buffer (by $Read$).
It is an important part of the specification of $Write$
but note that there is no corresponding rely condition in $Read$.
Firstly, there is the technical issue that $v$ is local to $Write$
and hence cannot be referred to in (the rely of) $Read$.
Secondly, several $Write$ operations might take place during a single $Read$ 
and hence there may be multiple changes to the buffer during a $Read$,
even though each $Write$ only changes the buffer (at most) once.
In fact, the possible multiple changes of the buffer during a $Read$
motivates the use of $\posvals{b}$ in its post condition.
It is worth observing that $\posvals{b}$ is applied to an abstract variable $b$ ---
the development that follows employs a representation of $b$ that is by no means obvious.

The guarantee of $Write$ requires that the observable effect of the operation
takes place in a single atomic step
and the use of the possible values notation in the post condition of $Read$ 
ensures that the observable effect of $Read$ also takes place in a single atomic step.
%That the observable effect of both operations takes place in a single atomic step
%links to Bornat's use of the concept of linearisability \cite{BornatAmjad13}. 

%\draftnote{An alternative view on $Write$ is $\atomicrel{\id}^\star ; \atomicrel{b' = v} ; \atomicrel{\id}^\star$
%and on $Read$, $\atomicrel{\id}^\star ; \atomicrel{r' = b} ; \atomicrel{\id}^\star$. Linearised specifications.
%}

The initial value of $b$ 
%(as in all specifications)
is assumed to contain a valid $Value$ so that it is possible 
for a $Read$ operation to precede the first $Write$.

Thus far, the possible values concept 
--that was devised in order to document an intermediate design-- 
has here been shown to offer a short and clear overall specification of ACM behaviour.
Freshness comes from the possible values notation and the effect of it being relative to the start of each $Read$ operation.
The implementation has to find a way of achieving the atomic change behaviour of $b$ in the abstraction without such an atomicity assumption.

\subsection{Understanding Simpson's representation}
\label{S-reif1}

The challenge of presenting a specification that makes sense to potential users
is addressed in Section~\ref{S-4-spec}.
A development using a single data reification step to a version of Simpson's code is presented in
Section~\ref{S-1step} ---
that development makes interesting further use of the possible values concept and is thus presented in some detail.
The current section attempts to provide an intuition of the `four-slot' data structure.
The operations corresponding to $Write$ and $Read$ of Section~\ref{S-4-spec} are here named
$Write\sb{i}$ and $Read\sb{i}$.

The importance of data abstraction and reification are commented on in Section~\ref{S-data-reification}.
Rather than jump immediately to Simpson's decision to use exactly four slots to represent the abstract variable $b$, a useful intermediate refinement step uses a data structure that contains
an abstract map of an indexed set of `slots'  $\mapof{X}{Value}$.
Here, this part of the state is named $dw$.
There is also a data type invariant that requires that the
(potentially partial) map has a value in every slot:
$\dom{dw} = X$.%
\footnote{Note that, in the concurrent context, the data type invariant must hold for every step,
not just initially and at the end of each operation.}

As in~\cite{JonesPierce10},
the index set $X$ is deliberately left unspecified at this stage.
$Write\sb{i}$ is decomposed into a three parts:

\begin{itemize}

\item $Write-ch\sb{i}$ chooses an index ($\in X$) that is safe to use

\item $Write-upd\sb{i}(v)$ updates the map $dw$ at the chosen index to $v$

\item $Write-com\sb{i}$ commits the index by exposing it to $Read\sb{i}$

\end{itemize}

\noindent
$Read\sb{i}$ is split into two parts:

\begin{itemize}

\item $Read-sel\sb{i}$ selects the most recently written index and stores that index in a local variable

\item $r \gets Read-acc\sb{i}$ accesses the indexed slot

\end{itemize}

$Write\sb{i}$ must inform $Read\sb{i}$ of the index of the slot which has been (most recently) written.
In addition, $Read\sb{i}$ must have a way of alerting $Write\sb{i}$ to the index of the slot that is claimed for reading.
Remember that the reader and writer processes are in no way synchronised and the implementation is designed to allow (multiple) reads to occur during a write or multiple writes to overlap with a single read.

It should be clear that the potential number of slots 
(the cardinality of the set $X$)
must be at least three because the writer has to select a member of $X$ that is neither the most recently written nor one which the reader might access 
(these could be the same but are not necessarily so).
It is possible to build a `three slot' implementation
{\em providing} there is an atomic way of communicating index values between $Read$ and $Write$.

%\cliffnote{CUT}

It is tempting to make $Read\sb{i}$ reserve a single element of $X$ to $Write\sb{i}$ but this does not actually provide an abstraction of Simpson's code.
What that code effectively does is to reserve more than one slot.
This is shown here as $pr$ being a set of indexes.

%\cliffnote{CUT}

The intermediate state is thus:

\begin{record}{\Sigma\sb{i}}
 dw: \mapof{X}{Value}  \mbox{\hspace{1em} -- space for values}\\
 lw: X \mbox{\hspace{4em} -- index of the last committed write} \\
 cw: X   \mbox{\hspace{4em} -- index claimed by the writer}\\% before it releases its chosen $X$ } \\
 cr: X \mbox{\hspace{4em} -- index claimed by the reader}\\
 pr: \setof{X} \mbox{\hspace{2.5em} -- potential elements of $X$ that the reader might use} 
 \end{record}
 
%\cliffnote{CUT}

It is an interesting observation that none of the variables can be modified by both operations.
The final letter of each variable name records which process,
\underline{r}eader or \underline{w}riter, can write to that variable
(e.g.~$lw$ can only be modified by $Write\sb{i}$).

It is not difficult to see the lines of the data reification required here:
the retrieve function is $b = dw(lw)$.
%
%\begin{fn}{retr}{mk-\Sigma\sb{i}(dw, lw, \cdots)}
%\signature{\Sigma\sb{i} \to Val}
%  dw(lw)
%\end{fn}
%
The initial state must, of course, satisfy the invariant;
the initial value in the buffer must be
$dw(lw)$ and there must be some arbitrary value in every slot
to ensure that $\dom{dw} = X$.

It is interesting to note that the issue of (data) race freedom on the slots is worked out 
at this level of abstraction
with rely/guarantee conditions.
This can be contrasted with Peter O'Hearn's view in~\cite{OHearn07} 
that separation logic is the tool of choice for reasoning about race freedom and rely/guarantee reasoning is for `racy' programs.
The decisive point appears to be that, here, race freedom is established on a data structure that is more abstract than the final representation.

%\cliffnote{CUT}

Although the observation is made above
that three slots would be adequate to avoid clashing,%
\footnote{In fact,~\cite{BornatAmjad13} also considers a three slot implementation.}
the genius of the representation proposed by Hugo Simpson is that 
--if four slots are used--
communication can be reduced to using single bits;
furthermore, 
in a physical implementation, 
these bits can be realised as wires connecting the $Read\sb{f}$ and $Write\sb{f}$ processes
running on separate processors.
Simpson describes the algorithm in terms of choosing `pairs' and `slots'.
As in~\cite{JonesPierce10},
this intuition is followed by using two sets $P$ and $S$ each of which has two possible values.
However, here,
toggling between the two values is achieved by a ``$\Not$'' operator.
Although both sets $P$ and $S$ can be implemented as Booleans,
the temptation to use Booleans is resisted at this stage
because separating the types $P$ and $S$ provides useful information
as to whether each index variable refers to a pair or a slot
(and has the potential to flag incorrect use as a type error).

%\cliffnote{CUT}

The final representation ($\Sigma\sb{f}$) is given in Figure~\ref{F-Sigma-f}.
%\cliffnote{Invariant: ??}
This is related to $\Sigma\sb{i}$ by a retrieve function where: 

\begin{itemize}

\item $dw$ is directly modelled by $dsw$ with the set $X$ reified to a $(P, S)$ pair

\item $cw$ is represented by $(cpw, \lnot sw(cpw))$

\item $cr$ is represented by $(cpr, csr)$

\item $lw$ is represented by $(lpw, sw(lpw))$

\item $pr$ is represented by $\set{(cpr, sl) | sl \in S}$

\end{itemize}

\begin{figure}
\begin{record}{\Sigma\sb{f}}
 dsw: \mapof{P \x S}{Value}  \mbox{\hspace{1em} -- two pairs of two data slots each} \\
 sw: \mapof{P}{S}  \mbox{\hspace{3em} -- $sw(p)$ is the last written slot for pair $p$} \\
 lpw: P   \mbox{\hspace{6em} -- last written pair} \\
 cpw: P   \mbox{\hspace{6em} -- current write pair} \\
 cpr: P   \mbox{\hspace{6em} -- current read pair} \\
 csr: S   \mbox{\hspace{6em} -- current read slot} 
\end{record}
\caption{The final state $\Sigma\sb{f}$}
\label{F-Sigma-f}
\end{figure}

\subsection{One step argument} %%%%%%%%%%%%%%%%%%%%
\label{S-1step}

This section presents a single-step data refinement from 
the top level specification using possible values to Simpson's algorithm.
Although the approach to refining the code from the specification is new,
the end code comes from Simpson's insights and motivates the approach.
The final state representation is as in Figure~\ref{F-Sigma-f}
and the relationship between the abstract buffer $b$ and this representation state is
\begin{eqnarray*}
  b = dsw(lpw,sw(lpw))\ .
\end{eqnarray*}
\begin{figure}
\begin{minipage}{0.43\textwidth}
\begin{formula}
Read\sb{f}()r: Value\T1
  \begin{array}{l}
        \Var t \in P ; \\
        \atomicassn{ t \gets lpw} ; \\
        \atomicassn{ cpr \gets t} ; \\
        \atomicassn{ csr \gets sw(cpr) }; \\
         r \gets dsw(cpr, csr)
  \end{array}
\end{formula}
\end{minipage}
\begin{minipage}{0.55\textwidth}
\begin{formula}
Write\sb{f}(v: Value)\T1
  \begin{array}{l}
           \atomicassn{ cpw \gets \Not cpr } ; \\
           dsw(cpw, \Not sw(cpw)) \gets v ; \\
           \atomicassn{ sw(cpw) \gets \Not sw(cpw) } ; \\
           \atomicassn{ lpw \gets cpw }
  \end{array}
\end{formula} 
\end{minipage}
\caption{Code for Simpson's algorithm}\label{f:Simpsons}
\end{figure}

The code for Simpson's algorithm is given in Figure~\ref{f:Simpsons}.
First note that the $Write\sb{f}$ operation has the following guarantee:%
\footnote{Although 
%(cf.~footnote~\ref{FN-cf}) 
ACMs are much more complicated than the one-place buffer,
the idea mentioned in Section~\ref{S-RG} of locating where a key value is unchanged without
adding auxiliary variables is evident here.}
\begin{eqnarray}
%  & (cpr,csr) = (cpw, \Not sw(cpw)) \implies dsw'(cpr,csr) = dsw(cpr,csr) 
%    \label{Write-guar-read-slot-stable} \\
  & \forall{i,j}{(i,j) \neq (cpw,\Not sw(cpw)) \implies dsw'(i,j) = dsw(i,j)} 
    \label{Write-guar-others-stable} 
%    \\
%  & cpw \neq cpr \implies cpw' \neq cpr'
%    \label{Write-guar-cpw-neq-cpr}
\end{eqnarray}
and that when the write operation is not active, 
the writing process does not modify any of the variables in the representation.

The specification of the $Read\sb{f}$ operation after mapping through the representation relation 
and extending the frame with an appropriate subset of the representation variables is
\begin{eqnarray}\label{dr-read-spec}
  \Spec{t,cpr,csr,r}{}{ r' \in \posvals{dsw(lpw,sw(lpw))} }
\end{eqnarray}
The first refinement step uses law~(\ref{law-posvals-pre-assignment}) from Section~\ref{S-sem}
with $lpw$ corresponding to $x$ and $dsw(v,sw(v))$ corresponding to $e$.
In fact the law needs to be extended to accommodate extra variables in the frame but that is straightforward.
\begin{eqnarray}
  (\ref{dr-read-spec}) & \refsto & \atomicassn{ t \gets lpw } ; \\
  && \Spec{cpr,csr,r}{}{ r' \in \posvals{dsw(t,sw(t))} } \label{second-spec}
\end{eqnarray}
For this one needs to rely on 
\begin{eqnarray}\label{rely-1}
lpw' \neq t \implies dsw'(t,sw'(t)) = dsw(t,sw(t))\ .
\end{eqnarray}
Note that $sw(t)$ is only changed by $Write\sb{f}$ if $t = cwp$ but the code also
guarantees that $t = cpw \implies t = lpw$ and hence $sw(t)$ can be changed only
if $lpw' = t$ and hence (\ref{rely-1}) holds.
If $sw(t)$ does not change, (\ref{rely-1}) is guaranteed by $Write\sb{f}$ 
by (\ref{Write-guar-others-stable}) because
$(t,sw(t)) \neq (cpw,\Not sw(cpw))$ because
either $t \neq cpw$ or if $t = cpw$ then $sw(t) = sw(cpw) \neq \Not sw(cpw)$.
This use of the slots vector $sw$ in Simpson's algorithm is one of the smart parts of how it works.

The second refinement step uses law~(\ref{law-posvals-pre-assignment}) once more to refine (\ref{second-spec}).
\begin{eqnarray}
  (\ref{second-spec}) & \refsto & \atomicassn{ cpr \gets t } ; \\
  && \Spec{csr,r}{}{ r' \in \posvals{dsw(cpr,sw(cpr))} } \label{third-spec}
\end{eqnarray}
provided one can rely on $t' \neq cpr \implies dsw'(cpr,sw'(cpr)) = dsw(cpr,sw(cpr))$
which holds trivially as $t = cpr$ is invariant over (\ref{third-spec}).

The third refinement step again uses law~(\ref{law-posvals-pre-assignment}) to refine (\ref{third-spec}).
\begin{eqnarray}
  (\ref{third-spec}) & \refsto & \atomicassn{ csr \gets sw(cpr) } ; \label{assign-csr} \\
  && \Spec{r}{}{ r' \in \posvals{dsw(cpr,csr))} } \label{fourth-spec}
\end{eqnarray}
provided one can rely on 
\begin{eqnarray}\label{third-rely}
  sw'(cpr) \neq csr \implies dsw'(cpr,csr) = dsw(cpr,csr)
\end{eqnarray}
being maintained by $Write\sb{f}$ for the duration of (\ref{fourth-spec}).
Here (\ref{third-rely}) can be strengthened to
\begin{eqnarray}\label{third-rely-rev}
  dsw'(cpr,csr) = dsw(cpr,csr)
\end{eqnarray}
and this can be shown to be maintained by $Write\sb{f}$ using the approach outlined
in Section~\ref{S-strengthening-law}.
For the duration of (\ref{fourth-spec}), $Read\sb{f}$ strengthens its guarantee to state
that it does not modify any shared variables ($r$ is local to $Read\sb{f}$).
Because (\ref{assign-csr}) establishes 
\begin{eqnarray}\label{read-phase-init}
 (cpr,csr) & \neq & (cpw,\lnot sw(cpw))
\end{eqnarray}
it is sufficient to show that $Write\sb{f}$ maintains (\ref{third-rely-rev}) from
initial states satisfying (\ref{read-phase-init}) provided there is no interference
on its shared variables.
If $Write\sb{f}$ is executing its write phase from any state satisfying (\ref{read-phase-init}),
it guarantees (\ref{third-rely-rev}) because the slot being written is not $(cpr,csr)$.
Once $Write\sb{f}$ finishes its write phase (or if it is not initially in its write phase)
it does not modify $dsw$ at all
(and hence maintains (\ref{third-rely-rev})) 
until after it executes $cpw \gets \lnot cpr$ 
which re-establishes (\ref{read-phase-init}) for the next write phase.

The final refinement step uses law~(\ref{law-single-reference-posvals}) to refine (\ref{fourth-spec}).
\begin{eqnarray*}
  (\ref{fourth-spec}) & \refsto & r \gets dsw(cpr,csr)
\end{eqnarray*}
This is valid provided $dsw(cpr,csr)$ is stable,
which follows from the argument given above for the previous refinement step.

A pleasing aspect of the above refinement is that, having started from a specification
(\ref{dr-read-spec}) using the possible values concept
(which allows for non-determinism in the value read),
the refinement steps have maintained the use of possible values 
(and hence the non-determinism) 
until the last step, when it is clear which slot is being read 
(and that the slot is stable).

Using the approach of locally strengthening a guarantee 
--and hence indirectly strengthening a rely-- 
(see Section~\ref{S-strengthening-law})
obviates the need to introduce and reason about auxiliary variables.
However, a development using auxiliary Boolean variables $reading$ and $writing$ is also possible,
where $reading$ is true if and only if the $Read$ process is actually reading from $dsw$
and $writing$ is true if and only if the $Write$ process is actually writing $dsw$.
With these auxiliary variables the important invariant is,
\begin{eqnarray*}
  reading \land writing & \implies &(cpr,csr) \neq (cpw,\lnot sw(cpw))
\end{eqnarray*}
which ensures that the $Read$ and $Write$ processes are not simultaneously
using the same slot.
This represents the weakest invariant to ensure correct operation of the algorithm. 

Because the specification of $Write\sb{f}$ does not make use of possible values notation 
its refinement is not presented in detail here.
An important property of $Write\sb{f}$ is that during its writing phase the slot being written 
differs from any slot that could be read concurrently,
which has been covered in the refinement of $Read\sb{f}$.
The other aspect of $Write\sb{f}$ worth noting is that its guarantee in the abstract
specification requires the buffer $b$ to be updated to $v$ atomically.
Recalling that $b$ is represented by $dsw(lpw,sw(lpw))$, 
that guarantee is achieved by (non-atomically) assigning to slot $(cpw,\lnot sw(cpw))$, 
which can never correspond to $b$.
The switch of $(lpw,sw(lpw))$ to $(cpw,\lnot sw(cpw))$ is then achieved 
either by the assignment $sw(cpw) \gets \lnot sw(cpw)$ if $lpw$ already equals $cpw$,
or by the following assignment $lpw \gets cpw$ if they differed initially.

Finally note that all of the atomic assignments in Figure\ \ref{f:Simpsons} are now
in a form in which they can be implemented by the corresponding non-atomic assignment,
assuming each read and write of a shared variable other than $dsw$ is atomic. 
This assumption is in line with the requirements because the shared flags can be implemented as single bits (or, indeed, realised as wires).

%\iannote{Done: Added ``other than $dsw$'' in last line above.}

\section{Conclusions and further work} %%%%%%%%%%%%%%%%%%%%%%%%%%%%%
\label{S-conc}

%\cliffnote{Cliff is still looking at this section\\
%but has moved subsection~\ref{s:auxiliary} here from Section~\ref{S-intro}}

The concept of possible values arose in an attempt to provide a clear design rationale of 
code which is delicate in the sense that slight changes destroy its correctness.
A seemingly simple and intuitive notational idea contributed to the description of a layered development.
The proposal was clearly motivated by a need in a practical application.
The next bonus came in the link to the non-deterministic state ideas:
this connection is set out in~\cite{HayesBurnsDongolJones12}.
The current paper contains the first publication of the specification given in Section~\ref{S-4-spec}
and the simplicity of the overall specification comes as strong encouragement
for the concept and notation of possible values.
This is further reinforced by the development of Simpson's algorithm in Section~\ref{S-1step}
which retains the use of the possible values notation and 
utilises laws taking advantage of the possible values notation.
%
%The authors recognise that this is the beginning of an exploration rather than a finished proposal
%but they hope that Jos\'{e} Nuno Oliveira will accept the tentative material as our birthday offering.

This closing section points to further avenues that appear to have potential but
certainly require more work.
As with the steps to date,
the motivation for the decisions should come from practical examples.

\subsection{Further applications} %%%%%%%%%%%%%%%%%%%%%%%%%%%%%

%A possible extension to the ACM requirements is to allow multiple reader processes.
%Although not in the original specification, 
%this extension has been looked at because it requires some thought in its formalisation.
%The specification of $Read$ in Section~\ref{S-4-spec} does not preclude concurrent reads
%but Simpson's 4-slot implementation does not handle this extension.

%In the spirit of the birthday conference where this paper was presented,
It can perhaps be mentioned that the possible values notation appears to have some potential 
for recording arguments about brain-teaser puzzles.
At the March 2015 meeting of IFIP WG~2.3 in Istanbul, 
Michael Jackson posed a hide-and-seek puzzle which is apparently described in several contexts.
Here, a mole is what must be located.
There are five holes in a line;
the mole moves each night to an adjacent hole;
the seeker can only check one hole per night and must devise a strategy that eventually locates the mole whose non-deterministic nocturnal movements are only constrained at either end of the line of holes.
This paper doesn't spoil the reader's fun by providing an answer;
it only mentions that one of the authors recorded the argument for termination using 
the possible values notation.

Sadly, most of the examples (see~\cite{SewellEt-11,vafeiadis-11,ridge-10})
using `weak memory' 
(a.k.a. `relaxed memory')
also give the feeling that they are gratuitous puzzles.
At a recent Schloss Dagstuhl meeting (15191),
one of the authors tried to use the possible values notation to record the 
non-determinism that results from not knowing when the various caches are flushed.
It must be conceded that, 
on the pure puzzle examples, 
possible values are doing little more than providing an alternative notation for
disjunctions.
A challenge is to find a genuinely useful piece of code that,
despite non-determinism,
satisfies a coherent overall specification under, say, 
total store order (TSO)
or
partial store order (PSO) memory models.
Only on such an application should the judgement about the usefulness of 
possible values be based.

%\iannote{The 4-slot algorithm should be OK with respect to total store order (TSO)
%but not partial store order (PSO).
%However, questions arise with TSO as to the meaning of posvals because the 
%writes of indices, in particular, may be delayed, 
%e.g. the updating of $lpw$ in memory may be delayed so that the reader gets
%an older value of $lpw$ (w.r.t. the writer's cache but not shared memory) 
%than at its start. 
%}

%\subsection{Options on the notation}\label{S-rg}%%%%%%%%%%%%%%%%%%%%%%%%%%%%%

There are also alternative views of the possible values notation itself.
For example, $\posvals{b}$ could yield a sequence of values rather than a set.
There is however an argument for preserving a (direct) way of denoting the set of possible values.

\subsection{Possible evaluations of expressions} %%%%%%%%%%%%%%%%%%%%%%%%%%%%%
\label{S-epe}

%\draftnote{But I would be in favour of retaining this link to rely conditions
%}

%\draftnote{cbj would like to drop the next piece:\\
%One possible use of possible values is to constrain the value of an output variable.
%For example, a post-condition with output variable $x$ of
%\[
%    \posvals{x} \subseteq [0..10]
%\]
%but such a condition could also be specified as an evolution invariant.
%}

As well as possible values of an expression $\posvals{e}$
that is the set of values of $e$ evaluated in each state of the execution,
one can define $\evalvals{e}$ as the set of all possible evaluations of $e$ over the execution interval:
each instance of a variable $x$ in $e$ takes on one of the values of $x$ in the interval
so that different occurrences of $x$ within $e$ may take on different values, and
the values of separate variables $x$ and $y$ may be taken from different states.
The set of evaluations includes those in which the values of all the variables are taken
in a single state and hence $\posvals{e} \subseteq \evalvals{e}$.
In~\cite{HayesBurnsDongolJones12} the possible values concept was linked to 
different forms of nondeterministic expression evaluation 
corresponding to $\posvals{e}$ and $\evalvals{e}$.

The following simple rule requires no restriction on $e$ 
other than it does not contain references to $x$ because $x$ is in the frame of the specification.
\[
  \rely{x' = x}{\Spec{x}{}{ x' \in \evalvals{e} } } \ \ \ \refsto\ \ \  x \gets e \ .
\]

If $e$ satisfies the single reference property over the execution interval 
(as defined earlier)
then $\evalvals{e} = \posvals{e}$ and hence
\[
 \begin{array}{ll}
     & \rely{x' = x \And (\bigwedge z \in S \suchthat z' = z)}{ \Spec{x}{}{ x' \in \posvals{e} } } \\
  = & \rely{x'=x \And (\bigwedge z \in S \suchthat z' = z)}{ \Spec{x}{}{ x' \in \evalvals{e} } } \\
  \refsto & x \gets e \ .
 \end{array}
\]

%\draftnote{
%The single reference property is rather syntactic -- can we come up with a more semantic approach?
%The semantic property is that evaluating the expression over an interval is equivalent to evaluating 
%it in the state in which the single reference variable is accessed.
%}
%
%\draftnote{Ian: I'd like to bring in rely conditions around here?\\
%to expand on this just a bit:\\
%to me, any tractable example is likely to have assumptions on how the environment can change the variables within $e$\\
%can we try to come up with an example?
%}

%Need to discuss the issue with (composite) values (such as arrays records but potentially integers (longs))
%for which read and write access isn't atomic.

%\draftnote{
%$\evalvals{e}$ can have the same implicit argument but the definition depends on the 
%form of the expression. For any expressions $e$ and $f$, and variable $x$,
%\begin{eqnarray}
%  \evalvals{e \oplus f}(t) & = & \{ v \in \evalvals{e}(t), w \in \evalvals{f}(t) \spot v \oplus w \} \\
%  \evalvals{\ominus e}(t) & = & \{ v \in \evalvals{e}(t) \spot \ominus v \} \\
%  \evalvals{x}(t) & = & \{ s \in \rng(t) \spot s(x) \}
%\end{eqnarray}
%where the latter assumes that accesses to variables are atomic;
%if this is not the case then a more complex definition is required
%that depends on the (atomicity) structure of $x$.
%}

\subsection{Auxiliary variables} %%%%%%%%%%%%%%%
\label{s:auxiliary}

The statement is made in~\cite{Jones-CARH-FS-10} that using auxiliary (a.k.a.\ ghost) variables in the specification of a software component {\em can} destroy compositionality by encoding too much information about the environment.
Studying possible values has helped put the position more clearly:

\begin{itemize}

\item having the code of the environment gives maximum information ---  but minimal compositionality

\item the same distinction is actually there with sequential programs where post conditions provide 
an abstract description of functionality without committing to an algorithm
(they can also leave unconstrained the values left in temporary variables etc.)

\item  for concurrency, things are much more sensitive:
one ideal is that the visible variables (read and write) of parallel processes are `separate' ---
this might be true on a concrete representation even when an abstract description appears to admit interference --- see~\cite{SEFM-15-paper}

\item rely/guarantee conditions are an attempt to state only what matters

\item the expressive `weakness' of rely/guarantee conditions (is conceded and) can be a positive attribute

\item auxiliary variables can be used to encode extra information about the environment ---
in the extreme, with use of statement counters, they can encode as much as the program
being executed by the environment

\end{itemize}

The advice is to minimise the use of auxiliary variables --- even when writing assertions,
abstraction from the environment can be lost if gratuitous information is recorded in auxiliary variables.
The `possible values' notation appears to offer an intuitive specification tool and 
a principled way of avoiding the need for some auxiliary variables.

%Add a short example?

One indication of the compositional nature of rely conditions is that, 
if a component with a rely condition $r$ is refined to a sequential composition,
each subcomponent inherits the rely condition $r$.
Conversely, a sequential composition guarantees a relation $g$
if each component of the sequential composition guarantees $g$.%
%\footnote{The general conditions to ensure such compositionality are explored more abstractly in \cite{AFfGRGRACP-TRX}.}

%\cliffnote{Ian is working on a one-step development\\
%using (logical) program counters, one can then write assertions of the form:\\
%$PC\sb{w} = \ldots \Implies \cdots$\\
%it also becomes possible to write
%$PC\sb{w} = \ldots \And PC\sb{r} = \ldots  \Implies \cdots$
%}

%\cliffnote{if this works, it throws up some interesting issues around ghost variables vs. abstraction\\
%the acknowledged `expressive weakness' of R/G 
%(which is also present with Owicki)
%can be circumvented by `ghost' variables (including program counters)\\
%but abstraction (sometimes?) comes to the rescue
%}

\section*{Acknowledgements} %%%%%%%%%%%%%%%%%%%%%%%%%%%%%

An earlier version of this paper was prepared for a conference that celebratied Jos\'{e} Nuno Oliveira's sixtieth birthday.
The authors of the current paper thank the organisers of that memorable event in Guimar\~{a}es and take the opportunity to renew their warmest good wishes to Jos\'{e}.

Outlines of material on possible values were presented at the 2015 meeting of IFIP WG~2.3
and at the Schloss Dagstuhl meeting 15191 ---
on both occasions comments were made that have helped clarify the ideas and their explanation.
In Dagstuhl, useful discussions with Viktor Vafeiadis helped one author understand 
the issues around weak memory.
Useful comments on a draft from Diego Machado Dias are gratefully acknowledged
as are those of the anonymous journal referees.

The research reported here is funded by
the EPSRC responsive mode grant on ``Taming Concurrency'', 
the EPSRC Platform Grant ``TrAmS-2''
and the ARC grant DP130102901; % on ``Understanding concurrent programs using rely/guarantee thinking''.
the authors express their thanks for this support.

\section*{References}

\bibliographystyle{alpha}
\bibliography{parallel}

%%%%%%%%%%%%%%%%%%%%%%%%%%%%%%%%
%%%%%%%%%%%%%%%%%%%%%%%%%%%%%%%%
\end{document}